\documentclass[twoside,reqno]{HERON}
\usepackage{epsfig,cite,colordvi}
\usepackage{graphicx}
\usepackage{url}
\usepackage{amssymb,amsmath,amscd,epsf}
\usepackage{times}
\usepackage{makeidx}
\newcommand{\np}{{\bf p}}

\newcommand{\Qbar}{\not{\!Q}}
\newcommand{\Kbar}{\not{\!K}}
\newcommand{\Pbar}{\not{\!P}}

\makeindex
\begin{document}

\title{Meson-exchange Currents and Quasielastic Neutrino Cross Sections}

\runningheads{Meson-exchange Currents and Quasielastic Neutrino Cross Sections}
{M.B. Barbaro et al.%
}

\begin{start}

\author{M.B. Barbaro}{1}, 
\coauthor{J.E. Amaro}{2}, 
\coauthor{J.A. Caballero}{3},
\coauthor{T.W. Donnelly}{4},
\coauthor{J.M. Ud\'ias}{5},
\coauthor{C.F. Williamson}{6}

\index{Barbaro, M.B.}
\index{Amaro, J.E.}
\index{Caballero, J.A.}
\index{Donnelly, T.W.}
\index{Ud\'ias, J.M.}
\index{Williamson, C.F.}

\address{Universit\`a di Torino, 10125 Turin, Italy}{1}
\address{Universidad de Granada, 18071 Granada, Spain}{2}
\address{Universidad de Sevilla, 41080 Sevilla, Spain}{3}
\address{Massachusetts Institute of Technology, Cambridge, MA 02139, USA}{4,6}
\address{Universidad Complutense de Madrid, 28040 Madrid, Spain}{5}

\begin{Abstract}
We illustrate and discuss the role of meson-exchange currents in quasielastic
neutrino-nucleus scattering induced by charged currents, comparing the 
results with the recent MiniBooNE data for differential and integrated cross 
sections.
\end{Abstract}
\end{start}

\section{Introduction}

The double differential quasielastic cross section for the charged-current 
quasielastic (CCQE) neutrino-nucleus process has been recently measured for 
the first time by the MiniBooNE collaboration at 
Fermilab~\cite{AguilarArevalo:2010zc}. 
Unexpectedly, the data turned out to be substantially underestimated by the 
relativistic Fermi gas (RFG) model used in the experimental analysis, 
as well as by several more realistic nuclear models. 
Indeed a phenomenological model based on electron scattering data, 
the super-scaling approach (SuSA)~\cite{Amaro:2004bs}, which describes 
by construction the world data on quasielastic electron scattering, yields 
cross sections which are lower than the RFG predictions and therefore in
worse agreement with the neutrino data.

This outcome has been initially ascribed~\cite{AguilarArevalo:2010zc} 
to an anomalously large value of the nucleon axial mass $M_A$, namely 
the cutoff parameter entering the dipole axial
form factor: in order to fit the data within the RFG model a value 
$M_A=1.35$ GeV/c$^2$ is required, significantly larger than the universally
accepted value $M_A\simeq 1$ GeV/c$^2$~\cite{Bernard:2001rs}. An even larger 
axial mass would be required in the SuSA model. Similar results are found
in the context of microscopical models such as the ones based on 
relativistic mean field 
theory~\cite{Caballero:2005sj,Amaro:2006tf} or realistic structure 
functions~\cite{Benhar:2010nx}, which have been widely tested against 
electron scattering.

However, as stressed in Ref.~\cite{Amaro:2004bs}, 
effects from meson exchange currents and their associated correlations 
are not accounted for in the SuSA approach, since they violate scaling of both 
kinds- that is, the 
corresponding superscaling function does depend on the momentum tranfer and
on the nuclear target, as shown in 
Refs.~\cite{Amaro:2002mj,Amaro:2003yd,DePace:2004cr}  - and were therefore 
ignored in analyzing the electron scattering data in terms of superscaling.
Although the effect of two-body currents in not very sizable at the quasielastic
peak in electron scattering, it can be more significant 
in quasielastic neutrino scattering due to the different kinematical conditions.
In fact in this case the neutrino beam is not monochromatic, but a wide energy 
range is spanned by the neutrino flux (from 0 to 3 GeV for MiniBooNE):
an event is classified as ``quasielastic'' if no pions are present in the 
final state, but it does not necessarily correspond to one-nucleon knockout.
In the calculations of Refs.~\cite{Martini:2009uj} and \cite{Nieves:2011pp} 
it has been shown that
multinucleon channels can account for the behavior of the CCQE cross sections 
without need of an anomalously large axial mass. On the other hand a different
model, based on the relativistic Green's function framework,
is also able to describe the experimental without the need to 
modify the nucleon axial mass, as recently shown on Ref.~\cite{Meucci:2011vd}.
Hence, before drawing any conclusion on 
the nucleon axial mass as extracted from neutrino data, a careful evaluation 
of all nuclear effects and of the relevance of multinucleon emission and of some
non-nucleonic contributions is required. 

A further point we would like to stress is that the kinematics of 
ongoing and future neutrino experiments
demands relativity as an essential ingredient and traditional 
non-relativistic models are questionable in this regime. 
Here we shall present a fully relativistic model for the meson exchange 
currents (MEC) associated to the pion and discuss the corresponding results for 
both electron and neutrino reactions. Further details and results can be
found in Refs.~\cite{Amaro:2010sd,Amaro:1998ta,Amaro:2010iu,Amaro:2002mj,Amaro:2003yd}.

\section{Meson Exchange Currents}

\begin{figure}[h]
\begin{minipage}{25pc}
\begin{center}
\includegraphics[width=18pc]{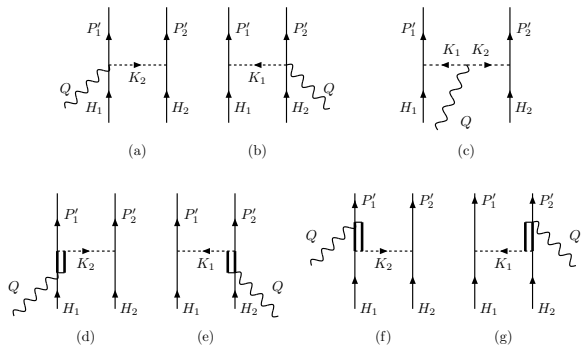}
\caption{ Two-body meson-exchange currents. 
(a) and (b): ``contact'', or ``seagull'' diagrams;  
(c): ``pion-in-flight'' diagram;
(d)-(g): ``$\Delta$-MEC diagrams 
(the thick lines represent the propagator of the $\Delta$-resonance).}
\label{fig:diagmec}
\end{center}
\end{minipage}
\end{figure}

Meson exchange currents are two-body currents carried by a virtual meson
exchanged between two nucleons in the nucleus.
The MEC considered in this work are represented by the Feynman
diagrams of Fig.~\ref{fig:diagmec}, where the dashed line represents a pion.

Assuming pseudo-vector nucleon-pion coupling, the fully relativistic
MEC matrix elements can be classified as follows~\cite{Amaro:2002mj,Amaro:2003yd}:
\begin{enumerate}
\item[1)]{Seagull or contact (diagrams $a$-$b$)}
\end{enumerate}
$$j^{\mu}_{s}
= \frac{f^2}{m_\pi^2}
             i\epsilon_{3ab}
             \overline{u}(\np'_1)\tau_a\gamma_5\Kbar_1 u(\np_1)
             \frac{F_1^V}{K_1^2-m_\pi^2}
             \overline{u}(\np'_2)\tau_b\gamma_5\gamma^{\mu}u(\np_2)
             + (1 \leftrightarrow 2) \,.
$$
\begin{enumerate}
\item[2)]{Pion-in-flight (diagram $c$)}
\end{enumerate}
$$j^{\mu}_{p}
=
\frac{f^2}{m_\pi^2}
             i\epsilon_{3ab}
             \frac{F_\pi(K_1-K_2)^\mu}{(K_1^2-m_\pi^2)(K_2^2-m_\pi^2)}
     \overline{u}(\np'_1)\tau_a\gamma_5\Kbar_1 u(\np_1)
             \overline{u}(\np'_2)\tau_b\gamma_5\Kbar_2 u(\np_2) \, .
$$
In the above 
$F_1^V$ and $F_\pi$ are
the electromagnetic isovector nucleon and pion form factors,
respectively and $f^2/4\pi = 0.08$ is the pion-nucleon coupling
constant.
\begin{enumerate}
\item[3)]{$\Delta$ current (diagrams $d$-$g$)}
\end{enumerate}
$$
j^{\mu}_{\Delta}
=
\frac{f_{\pi N\Delta} f}{m_\pi^2}
\frac{1}{K_2^2-m_\pi^2}
             \overline{u}(\np'_1)T_a^\mu(1) u(\np_1)
             \overline{u}(\np'_2)\tau_a\gamma_5\Kbar_2 u(\np_2)
             + (1 \leftrightarrow 2) \, .
$$
The vector $T_a^{\mu}(1)$ is related to the pion electroproduction
amplitude
$$
T_a^\mu(1)
=
K_{2\alpha}
\Theta^{\alpha\beta}
G^\Delta_{\beta\rho}(H_1+Q)
S_f^{\rho \mu}(H_1)
T_a T_3^{\dagger}
+
T_3 T_a^{\dagger}
S_b^{\mu \rho}(P'_1)
G^\Delta_{\rho\beta}(P'_1-Q)
\Theta^{\beta\alpha}
K_{2\alpha} \, 
$$
and involves the forward and backward $\Delta$ electroexcitation
tensors:
\begin{eqnarray}
S_f^{\rho \mu}(H_1)
&=&
\Theta^{\rho\mu}     \left(
       g_1  \Qbar
     - g_2 H_1\cdot Q
      +g_3Q^2
\right)
\gamma_5
-
 \Theta^{\rho\nu}Q_\nu
\left(
       g_1 \gamma^\mu
      -g_2 H_1^\mu
      + g_3Q^\mu
\right)
\gamma_5
\nonumber
\\
S_b^{\rho \mu}(P'_1)
&=&
\gamma_5
\left(
       g_1  \Qbar
     -g_2P'_1\cdot Q
     -g_3Q^2
\right)
\Theta^{\mu\rho}
-
\gamma_5
\left(
       g_1\gamma^\mu
      -g_2P'_1{}^{\mu}
      -g_3 Q^\mu
\right)
Q_\nu\Theta^{\nu\rho} \, ,
\nonumber
\end{eqnarray}
where
$g_i$ are the 
electromagnetic coupling constants,
$\Theta_{\mu\nu}=g_{\mu\nu}-\frac14\gamma_\mu \gamma_\nu $ and
$$ G^{\Delta}_{\beta\rho}(P)
=
 -\frac{ \Pbar+m_\Delta}{P^2-m_\Delta^2}
\left(
         g_{\beta\rho}
        - \frac{1}{3} \gamma_\beta\gamma_\rho
        - \frac{2}{3} \frac{P_\beta P_\rho}{m_\Delta^2}
        - \frac{\gamma_\beta P_\rho - \gamma_\rho P_\beta}{3m_\Delta}
\right) 
$$
is the Rarita-Schwinger $\Delta$ propagator.
Moreover we perform the substitution $m_\Delta\rightarrow
m_\Delta+\frac{i}{2}\Gamma(P)$ in the denominator of the propagator
to account for the $\Delta$ decay probability.
Our approach for the $\Delta$ follows, as a particular case, from
the more general form of the $\gamma N \Delta$ Lagrangian of
Pascalutsa {\it et al.}~\cite{Pas95}. 

\begin{figure}[h]
\label{fig:diagcor}
\begin{minipage}{25pc}
\begin{center}
\includegraphics[width=18pc]{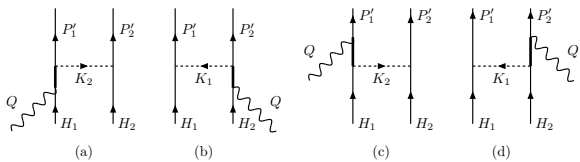}
\caption{Two-body correlation currents.}
\end{center}
\end{minipage}
\end{figure}

The MEC are not the only two-body operators able to induce 2p-2h
excitations. The correlation operators, arising from the Feynman diagrams
of Fig.~\ref{fig:diagcor}, are of the same order as the MEC in the
perturbative expansion and should be included in order to preserve the gauge
invariance of the theory. Their explicit expression can be found in 
Ref.~\cite{Amaro:2002mj}.  

In the next Subsections we shall illustrate the impact of these currents in
electron and neutrino scattering.

\subsection{Electron Scattering}

Meson exchange currents are carried by a virtual meson
which is exchanged between two nucleons in the nucleus.
Being two-body currents, the MEC can excite both one-particle one-hole 
(1p-1h) and two-particle two-hole (2p-2h) states.

In the 1p-1h sector, MEC studies of electromagnetic $(e,e^\prime)$ 
process have been performed for low-to-intermediate momentum transfers
(see, {\it e.g.}, \cite{Alberico:1989aja,Amaro:2002mj,Amaro:2003yd,Amaro:2009dd}),
showing a small reduction of the total response at the quasielastic peak,
mainly due to diagrams involving the electroexcitation of the
$\Delta$ resonance. 
However pionic correlation contributions, where the virtual boson is attached 
to one of the two interacting nucleons, have been shown to roughly compensate 
the pure MEC 
contribution~\cite{Alberico:1989aja,Amaro:2002mj,Amaro:2003yd,Amaro:2009dd}, 
so that in first approximation the contribution of two-body currents in the
1p-1h sector can be neglected.

In the 2p-2h sector, the contribution 
of pionic two-body currents to the electromagnetic response
was first calculated in the Fermi gas model in 
Refs.~\cite{Donnelly:1978xa,VanOrden:1980tg}, where sizable effects 
were found at large energy transfers.
In these references a non-relativistic reduction of the
currents was performed, while fully relativistic calculations have been
developed more recently in Refs.~\cite{Dekker:1994yc,DePace:2003xu,Amaro:2010iu}. 
In \cite{DePace:2003xu} only the pure MEC were considered, while in
\cite{Amaro:2010iu} the correlation diagrams were also included. The latter
present the problem of giving an infinite answer in a Fermi gas
model, due to a nucleon propagator that can be
on-shell in the region of the quasielastic peak and gives rise to
a double pole inside the integral.
Various prescriptions have been followed in order to avoid this 
problem~\cite{Alb84,Alb91,Gil97}, which is intrinsically related 
to the infinite extension of the Fermi gas. In Ref.~\cite{Amaro:2010iu}
we have dealt with the above divergence by means a regularization parameter 
$\epsilon$ which accounts for the finite size of the nucleus. An
exploratory study of the results has shown that a reasonable assumption for 
the regularization parameter, related to the propagation time of a real 
nucleon inside the nucleon, is $\epsilon\sim 200$ MeV, appreciably larger 
than the usual values of the nucleon width for collisions.

\begin{figure}[h]
\includegraphics[width=25pc]{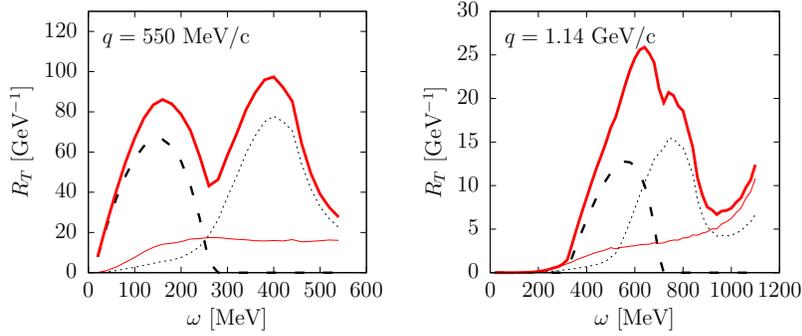}
\caption{\label{fig:Fe}(Color online)
Transverse response of $^{56}$Fe at $q=550$ and 1140
MeV/c. 
Dashed: RFG 1p-1h response with OB current only.
Dotted: MEC only. Thin solid (red): Correlation only for $\epsilon=200$ MeV. 
Thick solid (red): total one- plus two- body responses. 
} 
\end{figure}

In Fig.~\ref{fig:Fe} we show the transverse electromagnetic response 
function for the $^{56}$Fe for two values of the momentum transfer, 
$q=550$ and $1140$ MeV/c.
The contribution due to the full two-body current (MEC+correlations) 
in the 2p-2h sector (red, full solid) is compared with the 1p-1h response 
produced by the one-body current in the free relativistic Fermi gas (dashed). 
The separate contributions of the MEC (black, dotted) and correlations 
(red, thin solid) are also shown. 
It appears that the MEC produce 
a large peak with a maximum around $\omega= (m_\Delta^2+q^2)^{1/2}-m_N$, 
that comes from the $\Delta$ propagator appearing in the $\Delta$-current.
Indeed the latter turns out to dominate over the other diagrams, pion-in-flight
and seagull, and to be almost negligible in the longitudinal channel.
The presence of correlations leads to an additional, significant raise of the 
high energy tail. Moreover the correlation contribution, compared with the OB 
responses, is similar in the T and L channels, since its relative weight is 
independent of the particular component of the current 
(see Ref.~\cite{Amaro:2010iu}).

\subsection{Neutrino scattering}

In this section we apply the model above illustrated to CCQE neutrino 
scattering, implementing it in the phenomenological SuSA approach.

The CCQE neutrino-nucleus double differential cross section can be 
written according to a Rosenbluth-like decomposition as~\cite{Amaro:2004bs}
\begin{equation}
\left[ \frac{d^2\sigma}{dT_\mu d\cos\theta} \right]_{E_\nu} =
\sigma_0 \left[ {\hat V}_{L} R_L
+ {\hat V}_T R_T + {\hat V}_{T^\prime} R_{T^\prime} \right] ,
\label{eq:Ros}
\end{equation}
where $T_\mu$ and $\theta$ are the muon kinetic energy and scattering angle,
$E_\nu$ is the incident neutrino energy, $\sigma_0$ is the elementary 
cross section, ${\hat V}_i$ are kinematic 
factors and $R_i$ are the nuclear response functions, 
the indices $L,T,T^\prime$ referring to longitudinal, transverse,
transverse-axial, components of the nuclear current, respectively.
The response functions $R_L$ and $R_T$
have both ``VV'' and ``AA'' components 
(stemming from the product of two vector
or axial currents, respectively), whereas the axial response  $R_{T^\prime}$
arises from the interference of the axial and vector nuclear currents.

The SuSA approximation consists in modifying the well-known RFG response 
functions by replacing the free Fermi gas parabolic scaling function with
the phenomenological scaling function $f$ extracted from electron scattering
experimental data~\cite{Day:1990mf,Jourdan:1996ut}.
On the basis of the SuSA result, we have modified the nuclear responses 
according to the RFG predictions, as described in the previous section, 
to account for the effect of the MEC.
As previously explained, we can neglect in first approximation the
MEC in the 1p-1h sector and restrict our attention to 2p-2h final states.
Moreover, in lowest order the MEC affect only
the transverse polar vector response $R_T^{VV}$, since they are negligible in 
the longitudinal channel and suppressed in transverse axial channel.

\begin{figure}[h]
\includegraphics[width=9pc]{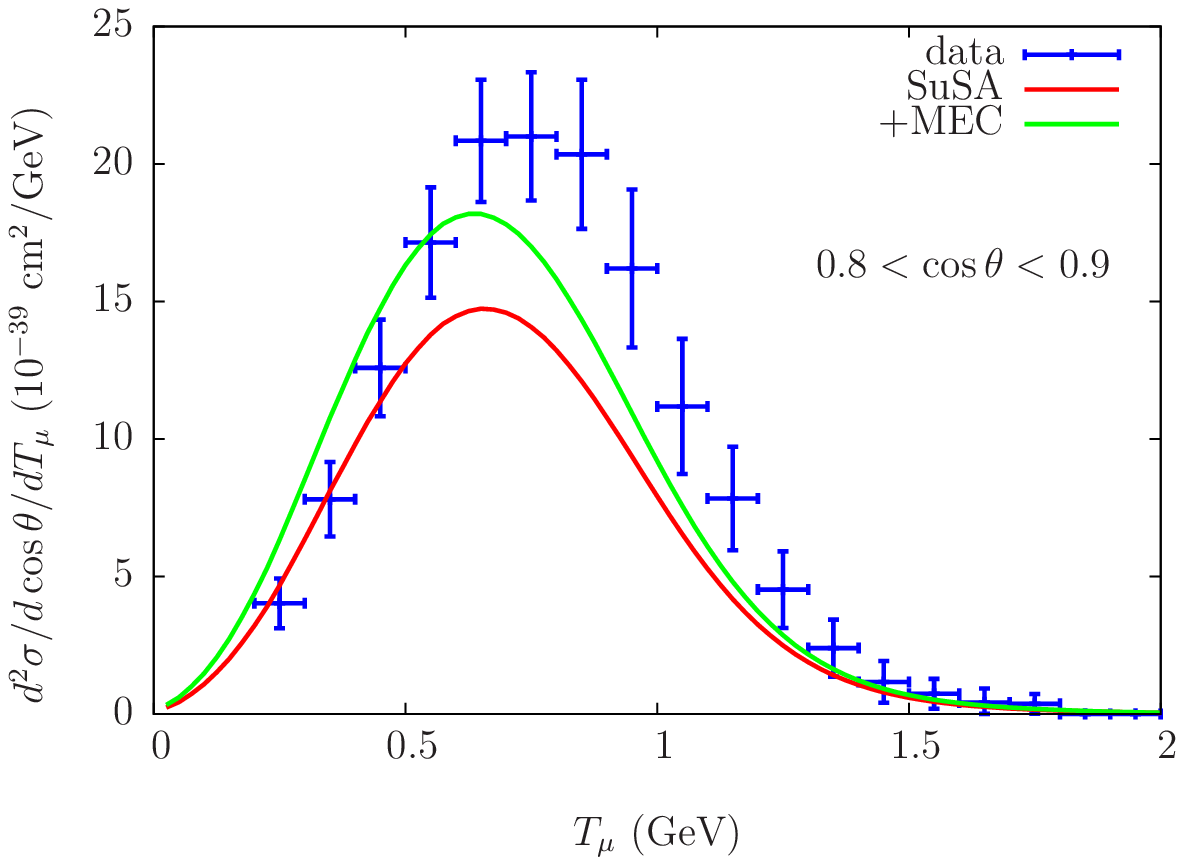}%
\includegraphics[width=9pc]{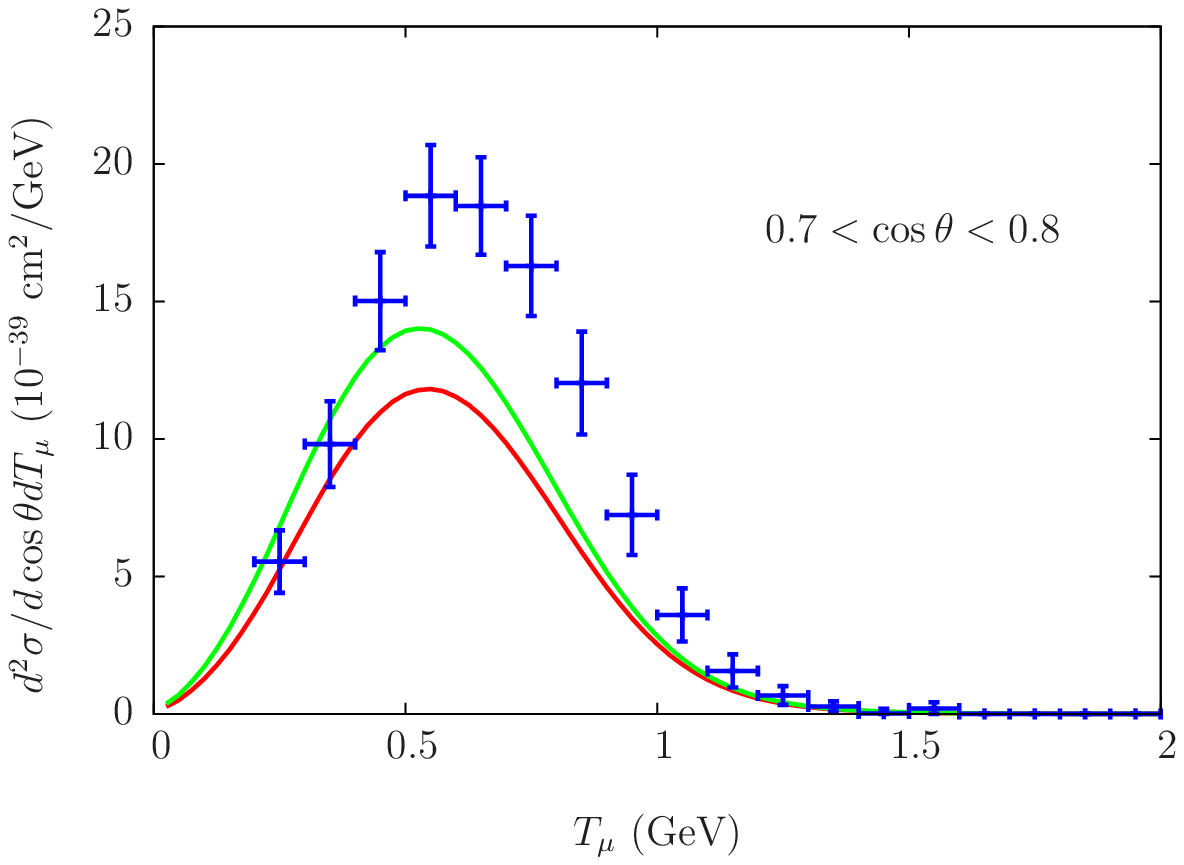}%
\includegraphics[width=9pc]{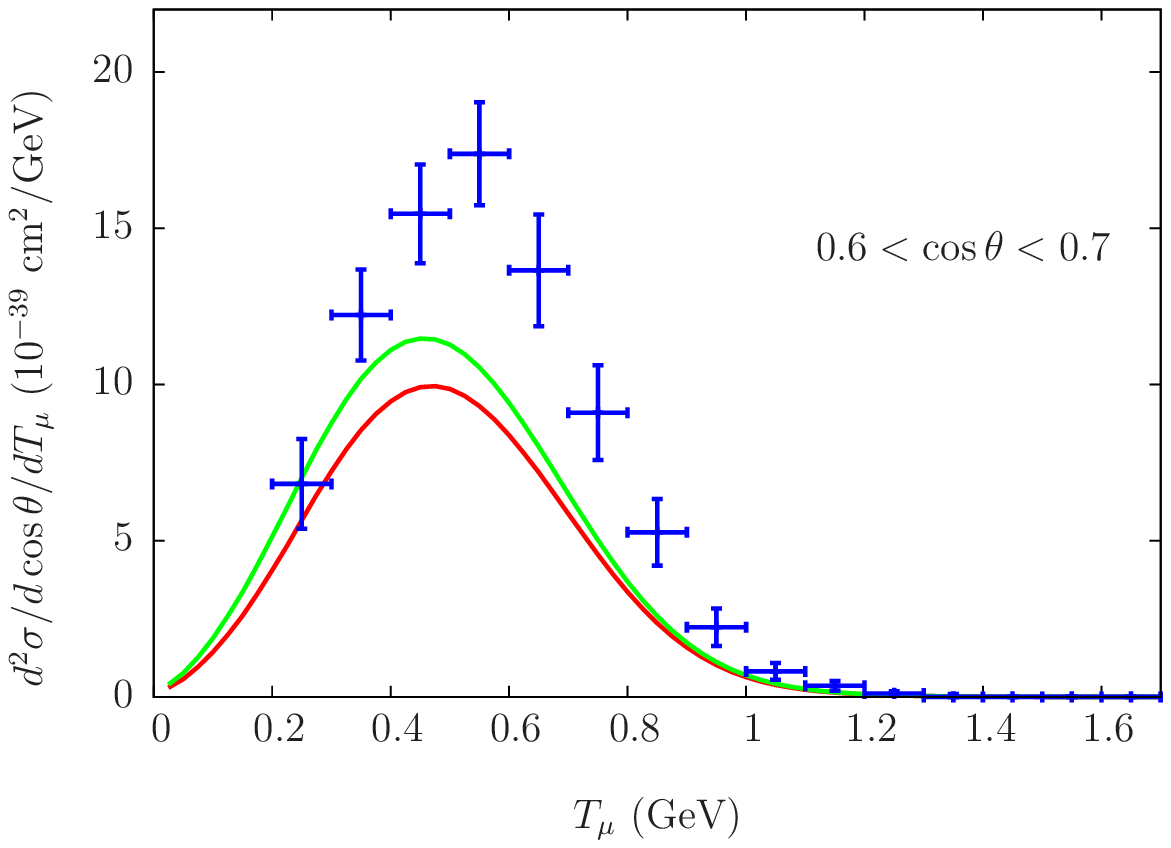}\\
\includegraphics[width=9pc]{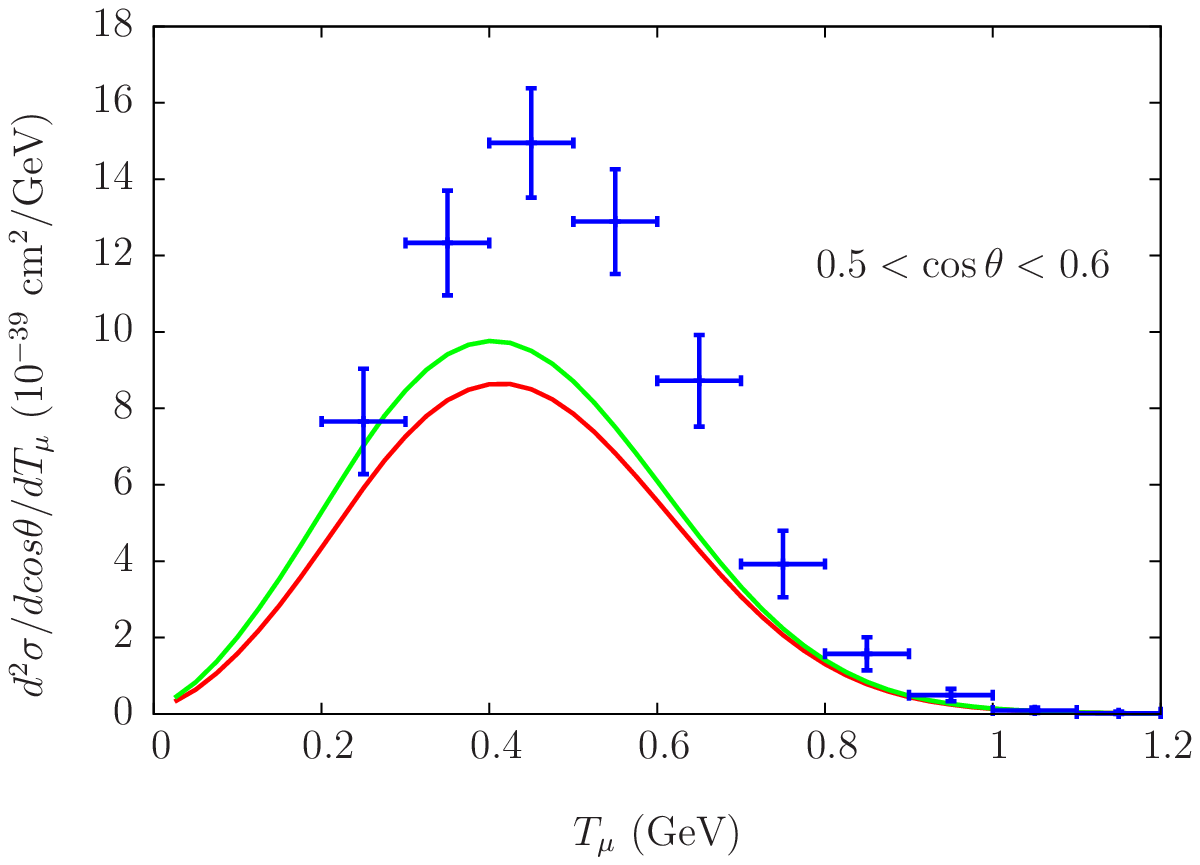}%
\includegraphics[width=9pc]{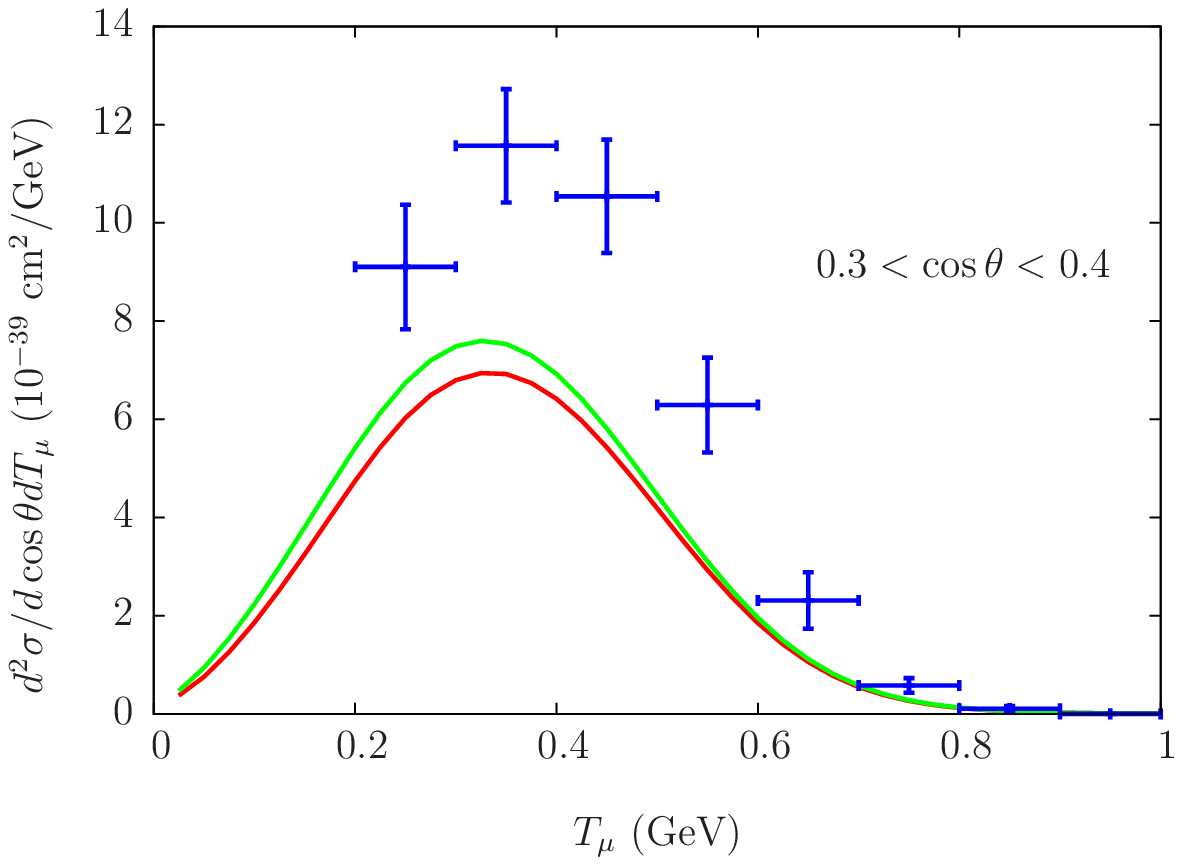}%
\includegraphics[width=9pc]{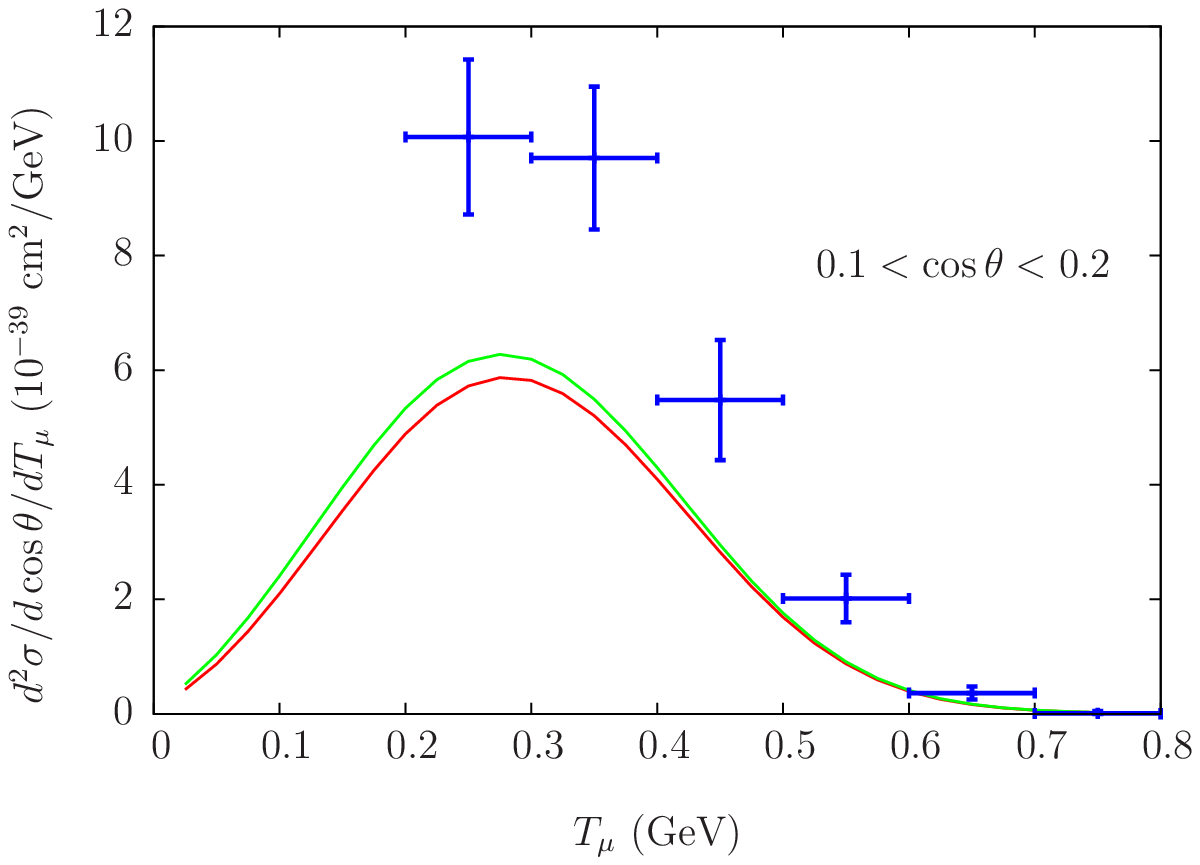}%
\caption{\label{fig:d2s} Flux-integrated $\nu_\mu$-$^{12}$C CCQE 
double differential cross section per target nucleon 
evaluated in the 
SuSA model with and without inclusion of 2p2h MEC displayed
versus the muon kinetic energy $T_\mu$ for various bins of
the muon scattering angle $\cos\theta$.
Here and in the following figures the data are from MiniBooNE~\cite{AguilarArevalo:2010zc}.}
\end{figure}

The corresponding results are shown in Fig.~\ref{fig:d2s}, where
the double differential CCQE cross sections obtained in the SuSA approach,
with and without inclusion of MEC, are compared with the MiniBooNE data after
averaging over the experimental neutrino flux.
It appears the 2p-2h MEC tend to increase the cross
section, yielding reasonable agreement with the data for not too high
scattering angles (up to $\cos\theta\simeq 0.6$). 
At larger angles the disagreement with
the experiment becomes more and more significant and the meson-exchange 
currents are not sufficient to account for the discrepancy. 

\begin{figure}[h]
\begin{minipage}{12pc}
\includegraphics[width=12pc]{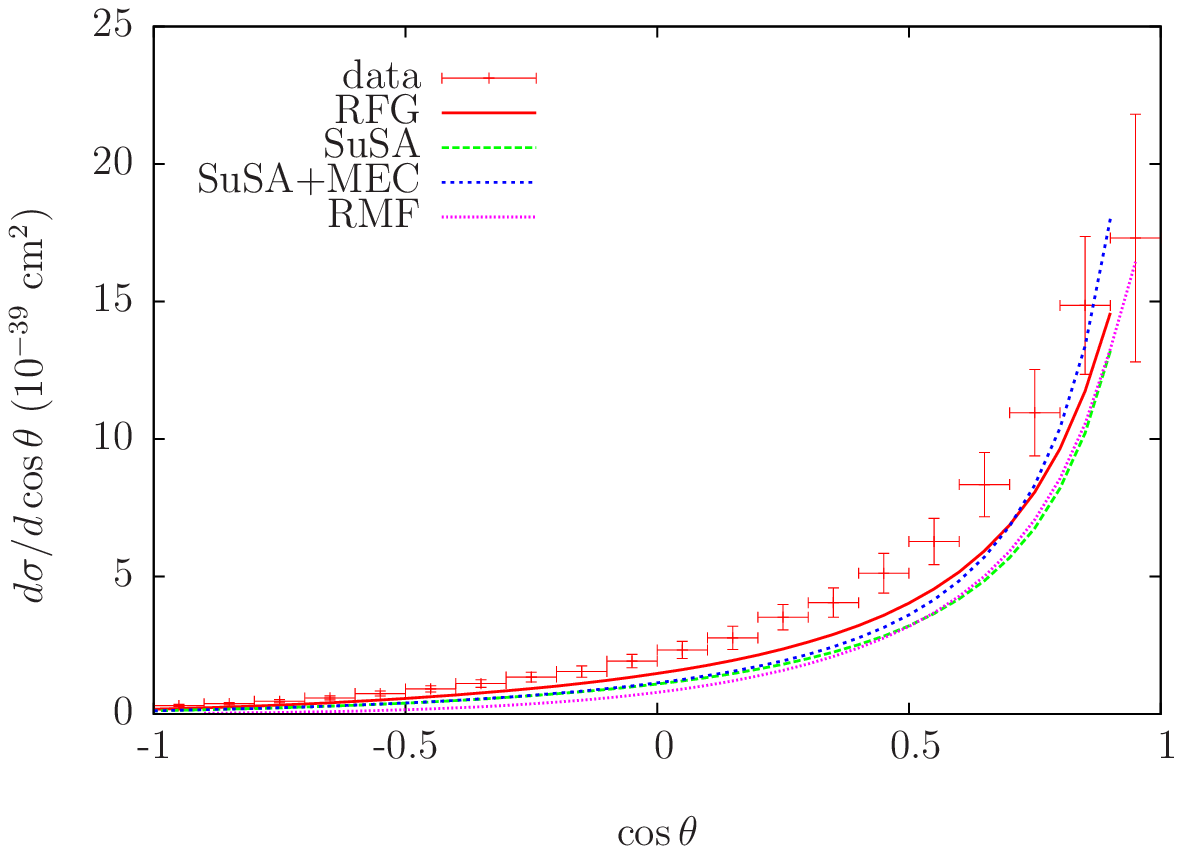}
\caption{\label{fig:dsdcos}(Color online) 
Flux-averaged cross section integrated over the scattering angle and
displayed versus the muon kinetic energy.}
\end{minipage}\hspace{2pc}%
\begin{minipage}{12pc}
\includegraphics[width=12pc]{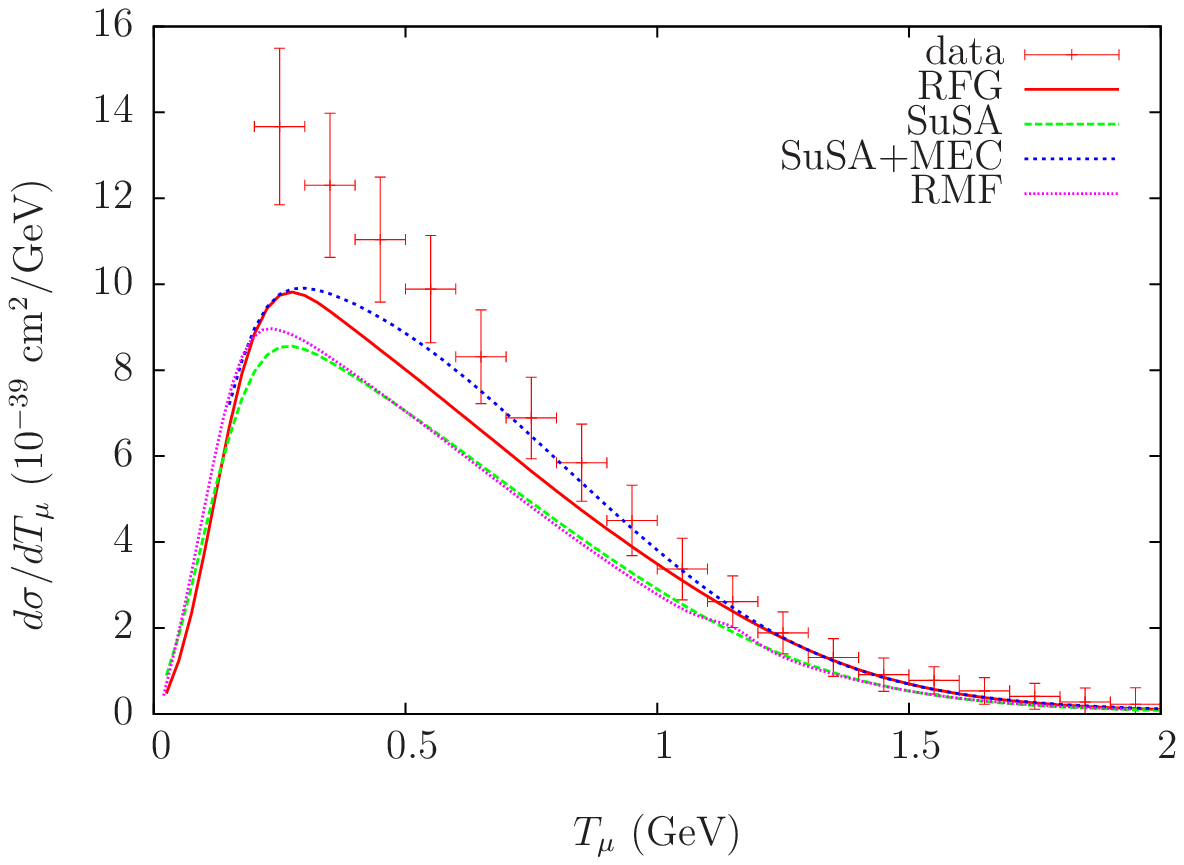}
\caption{\label{fig:dsdt}(Color online) 
Flux-averaged cross section integrated over the muon kinetic energy
and displayed versus the scattering angle.}
\end{minipage} 
\end{figure}

The single differential cross sections with respect to the muon kinetic
energy and scattering angle, respectively, are presented
in Figs.~\ref{fig:dsdcos} and \ref{fig:dsdt}, where the relativistic Fermi 
gas (RFG) and relativistic mean field (RMF) results are also shown for 
comparison:
again it appears the inclusion of 2p-2h excitations leads to a good agreement 
with the data at high $T_\mu$, but strength is still missing at the lower muon 
kinetic energies (namely higher energy transfers) and higher angles.

\begin{figure}[h]
\begin{minipage}{12pc}
\includegraphics[width=12pc]{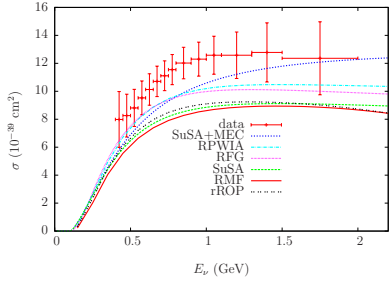}\hspace{2pc}%
\end{minipage}\hspace{2pc}%
\begin{minipage}{12pc}\caption{\label{fig:sigma}(Color online)
Total CCQE cross section 
per neutron versus the neutrino energy. The curves corresponding to different
nuclear models are compared with the flux unfolded
MiniBooNE data~\cite{AguilarArevalo:2010zc}.}
\end{minipage}
\end{figure}

Finally, in Fig.~\ref{fig:sigma} the fully integrated
CCQE cross section per neutron is displayed versus the neutrino energy 
and compared with the experimental
flux-unfolded data. Besides the models above discussed, we show
for comparison also the results of the relativistic mean field model 
when the final state interactions are ignored (denoted as RPWIA - relativistic
plane wave impulse approximation)
or described through a real optical potential (denoted as rROP).
Note that the discrepancies between the various models, observed in
Figs.~\ref{fig:dsdcos} and \ref{fig:dsdt},
tend to be washed out by the integration, yielding very similar
results for the models that include final state interactions (FSI)
(SuSA, RMF and rROP), all of 
them giving a lower total cross section than the models without FSI
(RFG and RPWIA). On the other hand the SuSA+MEC curve, while being
closer to the data at high neutrino energies, has a somewhat
different shape with respect to the other models, in qualitative
agreement with the calculation of \cite{Nieves:2011pp}.

Some caution should be expressed before drawing definitive conclusions 
from the agreements or disagreements seen in the results. For instance, 
there are strong indications from RMF studies as well as from QE $(e,e')$ 
data that the vector transverse response should be enhanced over the strict 
SuSA strategy employed here. 
Moreover, the correlation contributions, as shown in the previous section, 
are non-negligible in electron scattering when calculated in the RFG 
framework and should in principle be considered. 
However, besides the strong model dependence of these contributions,
associated to the already mentioned problem of the double pole, it is difficult 
to implement them in the SuSA model since some correlation effects may be 
already accounted for by the phenomenological scaling function, and simply 
summing the effects of RFG-based correlation diagrams to the SuSA responses 
would lead to double counting. Work is in progress to consistently include 
the correlation contribution in a microscopic relativistic model.

\begin{figure}[h]
\begin{minipage}{12pc}
\includegraphics[width=12pc]{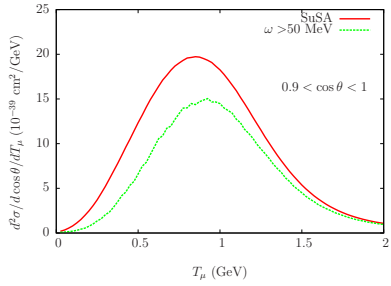}
\end{minipage}\hspace{2pc}%
\begin{minipage}{12pc}
\caption{\label{fig:cut}(Color online) Solid lines (red online): 
flux-integrated cross sections calculated in the SuSA 
model for a specific bin of scattering angle. Dashed lines (green online): 
a lower cut $\omega=50$ MeV is set in the integral over the neutrino flux.}
\end{minipage}\hspace{2pc}%
\end{figure}

A last comment is in order concerning the comparison with the data:
the average over the neutrino energy flux may require to
account for effects not included in models devised for quasi-free scattering. 
This is, for instance, the situation at the most forward scattering angles, 
where a significant contribution in the cross section comes from
very low-lying excitations in nuclei~\cite{Amaro:2010sd}. This is clearly 
illustrated in Fig.~\ref{fig:cut},
where the double differential cross section is evaluated in the SuSA model 
at the MiniBooNE kinematics and the lowest angular bin and compared
with the result obtained by excluding the energy transfers lower than 50 MeV
from the flux-integral. 
At these angles 30-40\% of the cross section corresponds
to very low energy transfers, where collective effects dominate and any
approach based on impulse approximation is inadequate to describe the 
nuclear dynamics.

\section{Conclusion}

In summary, we have shown that 2p-2h meson exchange currents play
an important role in CCQE neutrino scattering and may help to resolve
the controversy on the nucleon axial mass raised by the recent 
MiniBooNE data. 
In our approach two-body currents arise from
microscopic relativistic modeling performed for inclusive electron
scattering reactions and they are known to result in a significant
increase in the vector-vector transverse response function, in concert with 
QE electron scattering data. It should,
however, be remembered that the present approach, when applied to neutrino 
scattering, still lacks the contributions from the correlation diagrams 
associated with the MEC which are required by gauge invariance; 
these might improve the agreement with the data, as suggested by the results for
inclusive electron scattering.


\begin{thebibliography}{99}

%1
\bibitem{AguilarArevalo:2010zc}
  A.~A. Aguilar-Arevalo {\it et al.}  [MiniBooNE Collaboration],
  %``First Measurement of the Muon Neutrino Charged Current Quasielastic Double
  %Differential Cross Section,''
  \emph{Phys. Rev.}  \textbf{D81} 092005 (2010).
  %arXiv:1002.2680 [hep-ex].
  %%CITATION = ARXIV:1002.2680;%%
%2
\bibitem{Amaro:2004bs}
  J.~E. Amaro, M.~B. Barbaro, J.~A. Caballero, T.~W. Donnelly, A. Molinari and I. Sick,
  %``Using electron scattering superscaling to predict charge-changing  neutrino
  %cross sections in nuclei,''
  \emph{Phys. Rev.}  \textbf{C71} 015501 (2005).
  %%CITATION = PHRVA,C71,015501;%%
%3
%\cite{Bernard:2001rs}
\bibitem{Bernard:2001rs}
  V.~Bernard, L.~Elouadrhiri, U.~.G.~Meissner,
  %``Axial structure of the nucleon: Topical Review,''
 \emph{J. Phys.} \textbf{G28} R1-R35 (2002).
%  [hep-ph/0107088].
%4
\bibitem{Caballero:2005sj}
  J.~A.~Caballero, J.~E.~Amaro, M.~B.~Barbaro, T.~W.~Donnelly, C.~Maieron and J.~M.~Udias,
  %``Superscaling in charged current neutrino quasielastic scattering in the
  %relativistic impulse approximation,''
  \emph{Phys. Rev. Lett.}  \textbf{95} 252502 (2005).
%5
\bibitem{Amaro:2006tf}
  J.~E.~Amaro, M.~B.~Barbaro, J.~A.~Caballero and T.~W.~Donnelly,
  %``Quasielastic charged current neutrino nucleus scattering,''
  \emph{Phys. Rev. Lett.}  \textbf {98} 242501 (2007).
%6
%\cite{Benhar:2010nx}
\bibitem{Benhar:2010nx}
  O.~Benhar, P.~Coletti, D.~Meloni,
  %``Electroweak nuclear response in quasi-elastic regime,''
  \emph{Phys. Rev. Lett.}  \textbf{105} 132301 (2010).
%  [arXiv:1006.4783 [nucl-th]].
%7
\bibitem{Amaro:2002mj}
  J.~E.~Amaro, M.~B.~Barbaro, J.~A.~Caballero, T.~W.~Donnelly and A.~Molinari,
  %``Gauge and Lorentz invariant one-pion exchange currents in electron
  %scattering from a relativistic Fermi gas,''
  \emph{Phys. Rept.}  \textbf{368} 317 (2002).
%8
\bibitem{Amaro:2003yd}
 J.~E.~Amaro, M.~B.~Barbaro, J.~A.~Caballero, T.~W.~Donnelly and A.~Molinari,
  %``Delta-isobar relativistic meson-exchange currents in quasielastic  electron
  %scattering,''
  \emph{Nucl. Phys.}  \textbf{A723} 181 (2003). 
%9
%\cite{DePace:2004cr}
\bibitem{DePace:2004cr}
  A.~De Pace, M.~Nardi, W.~M.~Alberico, T.~W.~Donnelly, A.~Molinari,
  %``Role of 2p - 2h MEC excitations in superscaling,''
  \emph{Nucl. Phys.}  \textbf{A741} 249-269 (2004).
%  [nucl-th/0403023].
%10
\bibitem{Martini:2009uj} M.~Martini, M.~Ericson, G.~Chanfray, and J.~Marteau,
\emph{Phys. Rev.} \textbf {C80} 065501 (2009).
%
%\cite{Nieves:2011yp}
%\bibitem{Nieves:2011yp}
%  J.~Nieves, I.~R.~Simo, M.~J.~V.~Vacas,
  %``The nucleon axial mass and the MiniBooNE Quasielastic Neutrino-Nucleus Scattering problem,''
%11
%\cite{Nieves:2011pp}
\bibitem{Nieves:2011pp}
  J.~Nieves, I.~Ruiz Simo, M.~J.~Vicente Vacas,
  %``Inclusive Charged--Current Neutrino--Nucleus Reactions,''
\emph{Phys. Rev.}  \textbf{C83} 045501 (2011).
%  [arXiv:1102.2777 [hep-ph]].
%12
%\cite{Meucci:2011vd}
\bibitem{Meucci:2011vd}
  A.~Meucci, M.~B.~Barbaro, J.~A.~Caballero, C.~Giusti, J.~M.~Udias,
  %``Relativistic descriptions of final-state interactions in charged-current quasielastic neutrino-nucleus scattering at MiniBooNE kinematics,''
\emph{Phy. Rev. Lett.} \textbf{107} 172501 (2011).
%  [arXiv:1107.5145 [nucl-th]].
%13
\bibitem{Amaro:2010sd}
  J.~E. Amaro, M.~B. Barbaro, J.~A. Caballero, T.~W. Donnelly and C.~F. Williamson,
  %``Meson-exchange currents and quasielastic neutrino cross sections in the
  %SuperScaling Approximation model,''
  \emph{Phys. Lett.} \textbf{B696} 151 (2011). 
%  [arXiv:1010.1708 [nucl-th]].
  %%CITATION = PHLTA,B696,151;%%
%14
%\cite{Amaro:1998ta}
\bibitem{Amaro:1998ta}
  J.~E.~Amaro, M.~B.~Barbaro, J.~A.~Caballero, T.~W.~Donnelly and A.~Molinari,
  %``Relativistic effects in electromagnetic meson exchange currents,''
  \emph{Nucl. Phys.} \textbf{A643} 349 (1998)
%  [arXiv:nucl-th/9806014].
  %%CITATION = NUPHA,A643,349;%%
%15
%\cite{Amaro:2010iu}
\bibitem{Amaro:2010iu}
  J.~E.~Amaro, C.~Maieron, M.~B.~Barbaro, J.~A.~Caballero and T.~W.~Donnelly,
  %``Pionic correlations and meson-exchange currents in two-particle emission
  %induced by electron scattering,''
  \emph{Phys. Rev.}  \textbf{C82} 044601 (2010).
%  [arXiv:1008.0753 [nucl-th]].
  %%CITATION = PHRVA,C82,044601;%%
%16
\bibitem{Pas95} V. Pascalutsa and O. Scholten, 
\emph{Nucl. Phys.} \textbf{A 591} 658 (1995).
%17 
\bibitem{Alberico:1989aja}
  W.~M.~Alberico, T.~W.~Donnelly and A.~Molinari,
  %``PIONIC EFFECTS IN QUASIELASTIC ELECTRON SCATTERING,''
  \emph{Nucl. Phys.}  \textbf{A512} 541 (1998).
  %CITATION = NUPHA,A512,541;%%
%18
\bibitem{Amaro:2009dd}
  J.~E.~Amaro, M.~B.~Barbaro, J.~A.~Caballero, T.~W.~Donnelly, C.~Maieron and J.~M.~Udias,
  %``Meson-exchange currents and final-state interactions in quasielastic
  %electron scattering at high momentum transfers,''
  \emph{Phys. Rev.} \textbf{C81} 014606 (2010).
%  [arXiv:0906.5598 [nucl-th]].
%19
%\cite{Donnelly:1978xa}
\bibitem{Donnelly:1978xa}
  T.~W.~Donnelly, J.~W.~Van Orden, T.~De Forest, Jr., W.~C.~Hermans,
  %``Meson Exchange Currents In Deep Inelastic Electron Scattering From Nuclei,''
  Phys.\ Lett.\  {\bf B76}, 393 (1978).
%20
%\cite{VanOrden:1980tg}
\bibitem{VanOrden:1980tg}
  J.~W.~Van Orden, T.~W.~Donnelly,
  %``Mesonic Processes In Deep Inelastic Electron Scattering From Nuclei,''
  Annals Phys.\  {\bf 131}, 451-493 (1981).
%21
\bibitem{Dekker:1994yc}
M.J. Dekker, P.J. Brussaard, and J.A. Tjon,
Phys. Rev. C 49 (1994) 2650.
%22
\bibitem{DePace:2003xu}
  A.~De Pace, M.~Nardi, W.~M.~Alberico, T.~W.~Donnelly and A.~Molinari,
  %``The 2p-2h electromagnetic response in the quasielastic peak and beyond,''
  Nucl.\ Phys.\  A {\bf 726}, 303 (2003).
  %%CITATION = NUPHA,A726,303;%%
%23
\bibitem{Alb84}
W.~M.~Alberico, M.~Ericson and A.~Molinari,
\emph{Ann. Phys.} \textbf{154} 356 (1984).
%24
\bibitem{Alb91}
W.~M.~Alberico, A.~De Pace, A.~Drago and A.~Molinari,
\emph{Riv. Nuov. Cim.} \textbf{14} n.5, 1 (1991).
%25
\bibitem{Gil97}
A.~Gil, J.~Nieves and E.~Oset,
\emph{Nucl. Phys.} \textbf{A627} 543 (1997).
%26
\bibitem{Day:1990mf}
  D.~B.~Day, J.~S.~McCarthy, T.~W.~Donnelly and I.~Sick,
  %``Scaling in inclusive electron - nucleus scattering,''
  \emph{Ann. Rev. Nucl. Part. Sci.} \textbf {40} 357 (1990). 
%27
\bibitem{Jourdan:1996ut}
  J.~Jourdan,
  %``Quasielastic response functions: The Coulomb sum revisited,''
  \emph{Nucl. Phys.} \textbf{A603} 117 (1996).
%

\end{thebibliography}
\end{document}